\documentclass[aps, prl,twocolumn,amssymb,superscriptaddress, 10pt]{revtex4-1}

\usepackage{amsmath}    
\usepackage{graphicx}   
\usepackage{verbatim}   
\usepackage{xcolor}      
\usepackage{subfigure}  
\usepackage{hyperref}   
\usepackage{array}

\begin{document}

\title{Self-organizing Knotted Magnetic Structures in Plasma}
\author{C. B. Smiet}
\affiliation{Huygens-Kamerlingh Onnes Laboratory, Leiden University, P.O. box 9504, 2300 RA Leiden, The Netherlands}
\author{S. Candelaresi}
\affiliation{Division of Mathematics, University of Dundee, Dundee, DD1 4HN, UK}
\author{A. Thompson}
\affiliation{Department of Physics, University of California Santa Barbara, Santa Barbara, California, 93106, USA}
\author{J. Swearngin}
\affiliation{Department of Physics, University of California Santa Barbara, Santa Barbara, California, 93106, USA}
\author{J. W. Dalhuisen}
\affiliation{Huygens-Kamerlingh Onnes Laboratory, Leiden University, P.O. box 9504, 2300 RA Leiden, The Netherlands}
\author{D. Bouwmeester}
\affiliation{Huygens-Kamerlingh Onnes Laboratory, Leiden University, P.O. box 9504, 2300 RA Leiden, The Netherlands}
\affiliation{Department of Physics, University of California Santa Barbara, Santa Barbara, California, 93106, USA}

\begin{abstract}
We perform full-MHD simulations on various initially helical configurations and show that they reconfigure into a state where the magnetic field lines span nested toroidal surfaces. This relaxed configuration is not a Taylor state, as is often assumed for relaxing plasma, but a state where the Lorentz force is balanced by the hydrostatic pressure, which is lowest on the central ring of the nested tori. Furthermore, the structure is characterized by a spatially slowly varying rotational transform, which leads to the formation of a few magnetic islands at rational surfaces. We then obtain analytic expressions that approximate the global structure of the quasi-stable linked and knotted plasma configurations that emerge, using maps from $S^3$ to $S^2$ of which the Hopf fibration is a special case. The knotted plasma configurations have a highly localized magnetic energy density and retain their structure on time scales much longer than the Alfvenic time scale.

\end{abstract}

\maketitle

Understanding the types of structures in magnetic fields that occur in magnetohydrodynamics (MHD) is of fundamental importance for nuclear fusion \cite{White2001book, Hazeltine2003book} and astrophysics \cite{broderick2008fossil, Braithwaite2009, brandenburg2009critical,Tan2006}. Helicity-constrained, unbounded excitations in plasmas are present in a wide range of scales, from underdense bubbles emitted from active galactic nuclei ($\sim 10$ kpc) \cite{braithwaite2010magnetohydrodynamic}, through magnetic structures ejected from the solar corona ($\sim10^{5-6}$  km) \cite{low2001coronal} to the structure in fusion reactors such as the spheromak \cite{Jarboe1994} and field-reversed configurations \cite{Tuszewski1988} ($\sim$ m), and the plasmoids in dense plasma focus (DPF) experiments ($\sim$ mm) \cite{kubes2010transformation}. There exist many analytical solutions for the field in toroidal confinement vessels \cite{taylor1974} and bounded domains \cite{cantarella2001upper}, and even confinement vessels in the shape of a knot \cite{hudson2014new}. There are also analytical expressions for unbounded force-free fields \cite{marsh1996},  but no analytical expression has been found for a localized field that agrees with observed structures seen in unbounded plasmas. 

Magnetic helicity, defined as $H_{\rm m}=\int\mathbf{A}\cdot \mathbf{B} \ \mathrm{d}^3x$, where $\mathbf{A}$ and $\mathbf{B}$ are the vector potential respectively the magnetic field, was recognized by Woltjer to be an invariant of an ideal plasma \cite{Woltjer1958}. The identification of helicity as linking of magnetic field lines by Moffatt \cite{moffatt1969} gave a clear topological interpretation.  Given the topological nature of this invariant, Kamchatnov used the structure of the Hopf fibration to construct a topological soliton in ideal MHD \cite{Kamchatnov1982}. Recently this work was generalized by Thompson et al \cite{Thompson2014solitons}. This structure has not been described in resistive MHD, but also there helicity and magnetic topology play an important role in constraining magnetic relaxation \cite{taylor1974, berger1999introduction, DelSordo2010, ono1997experimental, Candelaresi2011helical}. In order to understand the effect of helicity in resistive plasmas we simulate the time evolution of various helical initial conditions  and find that each of them evolves towards an ordered state of nested toroidal magnetic surfaces.

\begin{figure*}
	\begin{center}
		\includegraphics[width=1.00\textwidth]{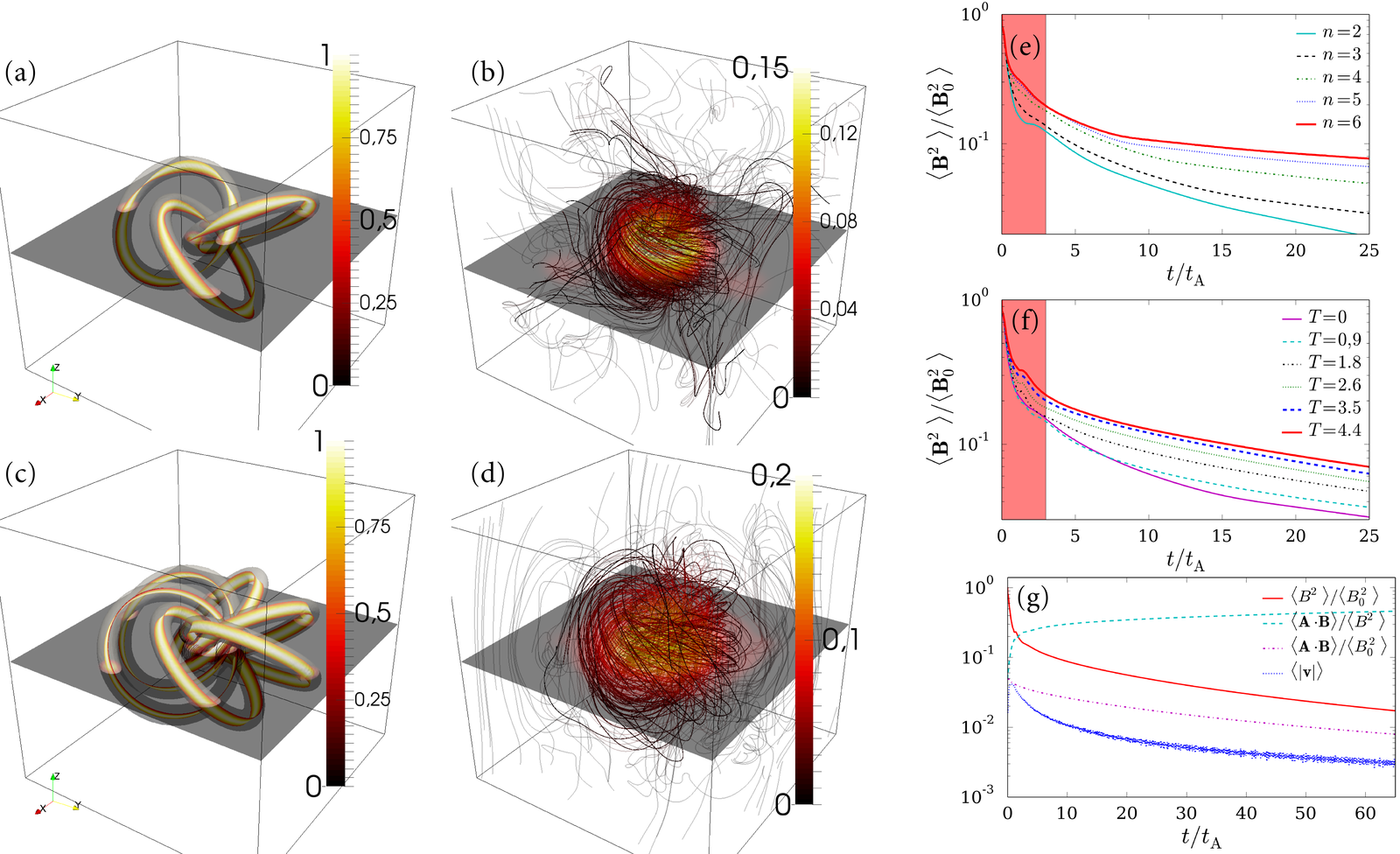}
	\end{center}
	\caption{Simulated configurations and evolution of magnetic fields. (a),(b) Initial condition and state at $t=22.5t_{\rm A}$ for the $n=3$ and $T=1.8$ simulation. (c)-(d) The same for the $n=6$ and $T=0$ simulation. In (a) and (c) a magnetic isosurface of $|\mathbf{B}|=0.1\mathcal{B}_0$ is shown to indicate the boundary of the flux tube. In (b), (d) the lines represent the magnetic field lines. The outer field lines are partially  transparent to not obstruct view of the central configuration. (e) Decay of magnetic energy for the simulations with $T=0$ and $n$ ranging from 1-6, and (f) the simulations with $n=3$ and $T$ ranging from 0-4.4. The shaded region indicates reconfiguration, after that resistive decay takes over. (g) The evolution of the average of magnetic energy $\langle B^2\rangle/\langle B_0^2 \rangle$, normalized helicity $\langle\mathbf{A}\cdot\mathbf{B}\rangle/\langle B^2\rangle$, helicity $\langle\mathbf{A}\cdot\mathbf{B}\rangle/\langle B^2_0\rangle$ and velocity $\langle|\mathbf{v}|\rangle$ for the simulation with $n=3$ and $T=1.8$.} 
	\label{fig1}
\end{figure*}

We simulate the plasma dynamics using the PENCIL CODE \cite{Brandenburg2002}. With this code we solve the resistive MHD equations in dimensionless form for an isothermal plasma in a fully periodic box of volume $(2\pi l_0)^3$ (see supplement). We choose as initial conditions simple configurations that are clear examples of fields containing
 helicity; rings of flux that are all linked and/or twisted. We start simulations with $n$ identical magnetic flux tubes that are all linked, with $n$ ranging from 2-6. The flux tubes have magnetic field of $1\mathcal{B}_0$ at the center of the tube, a radius of $\sqrt{2}l_0$, and a Gaussian intensity profile with characteristic width of $0.16l_0$. For the $n=3$ configuration we also vary the twist $T$ which indicates the number of windings of a field line around the center of the tube as it passes around the tube once, further increasing helicity. Two initial conditions are shown in figure \ref{fig1} (a) and (c). The velocity is initially zero everywhere and the density $\rho$ is set to 1 uniformly in the initial condition. The kinematic viscosity and magnetic diffusivity are $2\times 10^{-4}$, giving a magnetic Prandtl number of unity. The magnetic helicity of the initial conditions is given by $H_{\rm m} = (n^2-n)\Phi^2+nT\Phi^2$, where $\Phi$ is the magnetic flux through a single ring (see supplement).

The configurations evolve in a similar fashion, which can be divided into two regimes, reconnection and resistive decay, as shown in figure \ref{fig1}(e)-(f). We use Alfvenic time $t_{\rm A}=1/(2\sqrt{2}\pi l_0)$, scaled by the length of a flux tube. The tubes first contract, as this lowers the magnetic energy. This process is further detailed in the supplement. The higher the initial helicity, the less energy can be lost through reconfiguration. Figure \ref{fig1} (g) shows the evolution of several related quantities for the simulation with $n=3$ and $T=1.8$. 

\begin{figure}
	\centering
		\includegraphics[width=0.48\textwidth]{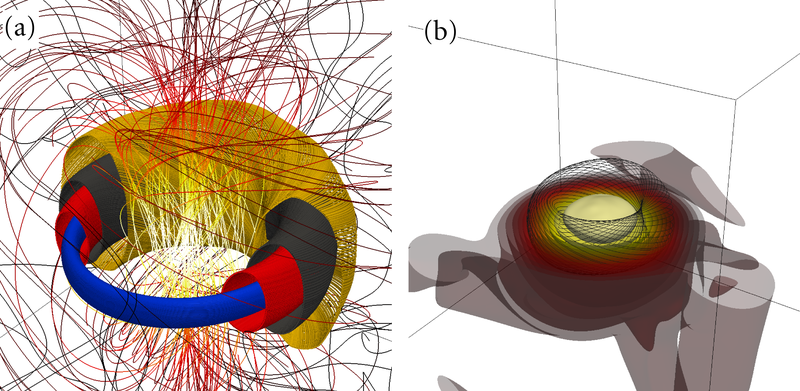}
	\caption{The simulation with $n=3$ and $T=1.8$ at time $t=22.5t_{\rm A}$. (a) The magnetic field contains invariant tori. Every surface is a single integral curve of the field of length $1000l_0$ colored differently for clarity. The surfaces are clipped to show the nested configuration. (b) Surfaces of constant magnetic field strength. A single torus is shown in black to indicate the scale of the magnetic structure.}
	\label{fig:finalfield}
\end{figure}

In order to analyze the emerging plasma configuration we take a detailed look at the simulation with $n=3$ and $T=1.8$ at time $t=22.5t_{\rm A}$. The magnetic energy is highly localized (figure \ref{fig:finalfield} (a)), falling off from the center. Remarkably, from the chaotic collapse of the initial condition, containing only a discrete rotational symmetry around the $z$-axis, an ordered magnetic structure emerges that is roughly axisymmetric and where field lines span invariant tori. These are toroidal surfaces spanned by magnetic field lines and are often described in the context of toroidal fusion devices \cite{hudson2012computation}. Four toroidal surfaces are shown in figure \ref{fig:finalfield} (a).

With higher initial helicity this structure appears sooner and is more pronounced. Invariant tori are observed in all simulations except the $n=2$ simulation which was stopped at $t=60t_{\rm A}$. In the $n=3$ and $T=0$ simulation tori were found only after $t=54t_{\rm A}$, but in all other simulations this structure appears before $t=22.5t_{\rm A}$ and remains. Invariant tori are also observed in simulations using different helical initial conditions, such as a single twisted ring and a trefoil knotted flux tube (see supplement). 

\begin{figure}
	\centering
		\includegraphics[width=0.48\textwidth]{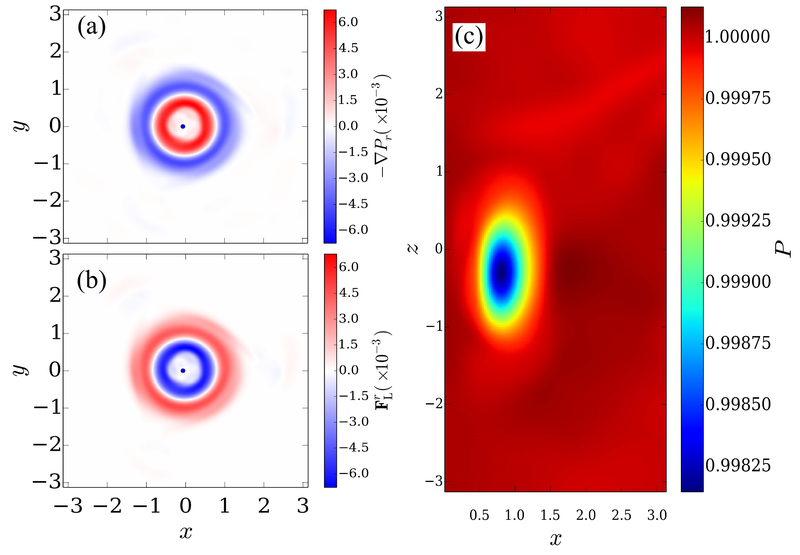}
	\caption{(a) The radial component of minus the gradient of the averaged pressure field, and (b) the radial component of the averaged Lorentz force, taken in the $x,y$-plane (top view). The geometrical center of the tori is taken as the origin , and is marked by the blue dot. (c) The pressure field in the $x,z$ half-plane, showing a lowered pressure in the center of the magnetic surfaces.}
	\label{fig:pressure}
\end{figure}

The initial reconfiguration of the rings induces pressure waves traveling through the periodic simulation volume. However, these pressure waves do not significantly affect the magnetic structure. To investigate the role of pressure in the simulation we average out these waves by averaging 365 snapshots between $t=27.5t_{\rm A}$ and $t=35.8t_{\rm A}$. Figure \ref{fig:pressure} (c) shows the averaged pressure, which is lowest on the magnetic axis of the structure. An ambient pressure $p_\infty$ is therefore inherent to the structure. The force due to the pressure gradient is balanced by the Lorentz force, which makes the structure quasi stable. In figure \ref{fig:pressure} (a) and (b) we show the average radial component of the Lorentz $F_{\rm L}^r$, and minus the average radial component of the pressure gradient $-\nabla P_r$ in the $x,y$-plane passing through the center of the structure (top view of the torus). 

Note that the \emph{lowered} pressure in the structure is consistent with the virial theorem \cite{bellan2006fundamentals, kulsrud2005plasma} that states that a free plasma cannot uphold an \emph{increase} in pressure solely by internal hydromagnetic forces. The region of highest magnetic field strength is near the geometrical center of the tori, where the pressure is unchanged. 

The balance of magnetic and hydrostatic forces indicates that the magnetic field forms self-stable, localized structures in MHD equilibrium with ambient pressure $p_\infty$. These equilibria feature rich dynamics such as the formation of magnetic islands at rational surfaces that are an area of intense research \cite{hudson2012computation, fitzpatrick2011fundamentals}. 

\begin{figure}
	\centering
		\includegraphics[width=0.48\textwidth]{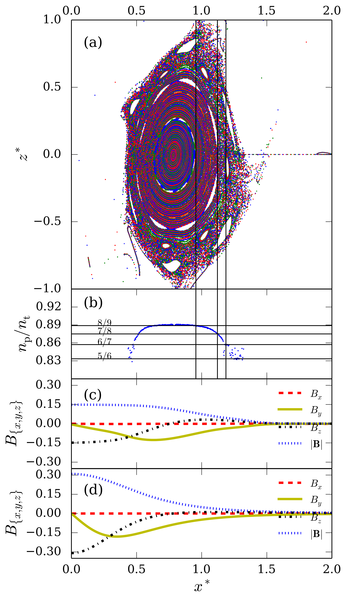}
	\caption{Poincar\'e plot and properties of the force-balanced toroidal structure. (a) Poincar\'e plot of the magnetic field. (b) The value of the rotational transform for each magnetic surface. Rational values are indicated by the labeled horizontal lines, and the positions where they cross are shown by vertical lines. (c) Value of the magnetic field strength, and the components at each position. (d) The magnetic field strength and components of the magnetic field for the analytical expression of a field with the same energy and rotational transform.} 
	\label{poincare}
\end{figure}

In order to investigate the nature of this equilibrium we construct a Poincar\'e plot of the field in figure \ref{fig:finalfield}. As seed points we choose 500 points on a line from the geometrical center of the tori  and through the  magnetic axis, starting on the magnetic axis and moving outward. We label this direction $x^*$, and the direction perpendicular to that and out of the plane of the torus we call $z^*$. The field lines are traced for a distance of $4000l_0$, and the positions where they cross the plane defined by $x^*$ and $z^*$ are marked. The Poincar\'e plot is shown in figure \ref{poincare} (a). We show the rotational transform $\imath$ \cite{goedbloed2004principles} of the corresponding field line in figure \ref{poincare} (b) (see supplement for calculation).

As expected from \cite{hudson2012computation}, where the rotational transform crosses rational values we observe magnetic islands. Lines are drawn indicating where the rotational transform crosses the values $8/9$, $7/8$, $6/7$ and $5/6$. As expected the number of islands observed is equal to the denominator of the rotational transform. Even though $\imath$ crosses a few rational surfaces, the value varies less than 10\%. In a tokamak equilibrium, where the inverse of $\imath$, the safety factor $q$, is used, this value typically varies much more \cite{goedbloed2010advanced}. We note that the fact that our pressure plots result from an averaging over time implies that we cannot resolve the fine structure in the pressure, such as possible discontinuities in pressure over specific irrational KAM surfaces as described in \cite{mcgann2010hamilton}. 

The magnetic field strength and the $x^*$, $y^*$, and $z^*$-components (with $y^*$ perpendicular to $z^*$ and $x^*$) at the position of the seed points are shown in figure \ref{poincare}. The field varies continuously over the surfaces, and the magnitude is higest in the geometrical center of the structure.

The part of the magnetic field that forms toroidal magnetic surfaces is reasonably axisymmetric, and could in principle be approached by a solution to the Grad Shafranov equation \cite{kulsrud2005plasma}. This would however not capture the large part of the field outside of this ordered region. Instead we want to point out a curious resemblance between the structure of the Hopf fibrations and the fields observed here. 

Non-null-homotopic maps (functions) from $S^3$ (hypersphere) to $S^2$ (sphere) such as the Hopf map \cite{Hopf1931} feature a topolocical structure resembling the observed plasma sctructures. The fibers (pre-images of points on $S^2$) of the map are continuous curves that lie on the surfaces of nested tori. Furthermore, every fiber is linked with every other one, with linking number depending on the map. Through stereographic projection from $S^3$ to $\mathbb{R}^3$ the fiber structure of this map can be translated to a vector field in $\mathbb{R}^3$ whose integral curves are the fibers of this map, (derivation in the supplement). Moreover, the obtained field is smooth, continuous, divergenceless, has helicity, and the field lines lie on the surfaces of nested tori. 

This curious structure was used by Kamchatnov to describe a soliton in ideal MHD, where the fluid velocity is parallel to the magnetic field everywhere \cite{Kamchatnov1982}. Independently, Ra\~nada, used the structure of the Hopf map to construct full radiative solutions of Maxwell's equations \cite{Ranada1989, Ranada1995}. Kamchatnov's solution was generalized by Sagdeev \cite{Sagdeev1986nonlinear}, and a similar extension of Ra\~nada's fields was described by Array\'as and Trueba \cite{arrayas2011electromagnetic}. 

The analytical form of this vector field in $\mathbb{R}^3$ is given by:
\begin{equation}
\mathbf{B} = \frac{4r_0^4\sqrt{a}}{\pi(r_0^2 + r^2)^3}
\begin{pmatrix} 2(\omega_2 r_0 y- \omega_1 xz  ) \\ -2( \omega_2 r_0 x + \omega_1 yz) \\ \omega_1(-r_0^2+x^2+y^2-z^2)  \end{pmatrix}.
\label{Bfield}
\end{equation}
This field is cylindrically symmetric around the z-axis (see supplement). It has a finite magnetic energy, as can bee seen from:
\begin{equation}\label{energy}
\int B^2 \ \mathrm{d}^3 x= a r_0^3(\omega_1^2+\omega_2^2),
\end{equation}
and nonzero helicity, given by:
\begin{equation}\label{saghel}
H_{\rm m}=a r_0^4\omega_1\omega_2,
\end{equation}
and like the field in the simulation it tends to zero away from the center. 

In our observed structure, the fluid velocity is neither parallel nor proportional to the magnetic field, making this structure fundamentally different from the structure Kamchatnov described \cite{Kamchatnov1982}. Nevertheless, the magnetic topology of field lines lying on nested toroidal surfaces, the magnetic energy localized in the center, the near constant rotational transform, and the direction of the magnetic field, even outside of the area that forms magnetic surfaces, all agree qualitatively with the toroidal structure described by equation \ref{Bfield}. To quantify this claim we extract from the simulation the parameters $\omega_1$, $\omega_2$, and $r_0$, needed for equation \ref{Bfield} (method described in the supplement), and show that there is overall agreement.  

For the simulation with $n=3$ and $T=1.8$ this yields values of $r_0=0.78$, $\omega_1=0.24$ and $\omega_2=0.27$. Parameters for the other simulations are quite similar (see supplement). The analytical magnetic field is shown in figure \ref{poincare} (d) for the same positions as the extracted field in (c). Even though there are differences in the magnitude of the components, there is broad agreement, which is quite remarkable for a routine that only uses three independent variables that are not fit, but calculated from select parameters extracted from the simulation. As time elapses $r_0$ increases and $\omega_1/\omega_2$ decreases. This change is such that over $45t_{\rm A}$ $r_0$ increases by 35\% and $\omega_1/\omega_2$ decreases by 50\%.


We have shown that reconnection of helical fields in resistive MHD causes the emergence of a self-stable toroidal magnetic field in force equilibrium. This  equilibrium results from a balancing of magnetic forces and the pressure gradient, and has a minimum in pressure on the magnetic axis. Note that this is not a Taylor state, and the pressure profile is inverse to the pressure enforced in a Tokamak reactor. In the quasistable state there is rich dynamics such as the emergence of magnetic islands at rational surfaces. 

Furthermore, we obtained an analytic expression for a magnetic field whose field lines lie on nested tori, requiring only three independent parameters. This field is a good approximation for the plasma configurations that emerge in the simulations, where a significant portion of the magnetic field lines reconfigure to lie on nested toroidal magnetic surfaces. We have observed the formation of this self-stable structure for various initial plasma configurations containing helicity. This indicates that this structure is a fundamental self-confining configuration that we predict to occur in situations where there is unbounded plasma containing helicity. 
 
\begin{acknowledgments}
We wish to thank Marco de Baar, Hugo de Blank, Hans Goedbloed, Hennie van der Meiden and Egbert Westerhof of the FOM DIFFER institute for stimulating discussions on plasma confinement.
This work is funded by NWO VICI 680-47-604 and NSF Award PHY-1206118. CBS was supported by the NWO Graduate Programme. SC acknowledges financial support from the UK’s STFC (grant number STK/K000993/1). We acknowledge support from the Casimir Research program Leiden-Delft and from NWO through a Spinoza award. 
\end{acknowledgments}

\newpage
\section{APPENDIX: supplemental material}

\section{Magnetic Helicity}
We calculate the helicity of the initial condition consisting of $n$ linked rings each with a twist $T$, each carrying the same magnetic flux $\Phi$. This calculation holds for thin rings. The helicity integral 
\begin{equation}
H_{\rm m}=\int_V\mathbf{A}\cdot\mathbf{B}\ \mathrm{d}^3x
\end{equation}
can be split into the contributions of every single ring separately, since the magnetic field is zero outside the flux rings. For a thin ring the integral can be approximated by
\begin{equation}\label{helicityint}
H_{\rm m}^{(1)}=\Phi \oint \mathbf{A} \cdot \ \mathrm{d}\mathbf{l}.
\end{equation}
Here $\Phi$ is the magnetic flux in a single ring and the integral is evaluated along a closed curve running along the center of the flux ring. Through Stokes' theorem the integral can be rewritten to:
\begin{equation}
H_{\rm m}^{(1)}=\Phi\iint\limits_{S_2}\nabla\times\mathbf{A} \cdot \ \mathrm{d}\mathbf{a} =  \Phi \left(\Phi_\text{encl}\right),
\end{equation}
where the definition for the vector potential $\nabla\times\mathbf{A}=\mathbf{B}$ has been used, and $\Phi_{\text{encl}}$ indicates the magnetic flux throught the surface enclosed by the curve. This has two contributions
\begin{equation}
\Phi_{\text{encl}}=\Phi_{\text{self}} + \Phi_{\text{others}},
\end{equation} 
where $\Phi_\text{self}$ indicates the contribution of flux by the field of the ring itself, caused by twist in the ring, and $\Phi_\text{others}$ indicates flux caused by other flux tubes passing through the center of the ring integrated over. 

The contribution of $\Phi_\text{others}$ is proportional to the number of rings passing through the ring in question, and is equal to $\Phi_\text{others}=(n-1)\Phi.$ The contribution $\Phi_\text{self}$ of self-linking in a twisted flux tube is proportional to the twist; $\Phi_\text{self}=T\Phi.$ This defines the helicity of a single ring, the helicity of the entire configuration is simply found by multiplying by $n$:
\begin{equation}
H_{\rm m}=(n^2-n)\Phi^2+nT\Phi^2.
\end{equation}

The parameters $n$ and $T$ increase the amount of magnetic helicity in the initial condition. For thin rings $T$ only changes the magnetic helicity of the initial configuration, not the magnetic energy.

\section{Equations solved by the PENCIL-CODE}

The PENCIL-CODE is used to solve the resistive magnetohydrodynamical (MHD) equations for an isothermal plasma in sixth order finite differences in space and third order in time. We solve the equations in a fully periodic box of size $(2\pi l_0)^{3}$ with $256^{3}$ meshpoints. We solve for the vector potential $\mathbf{A}$ instead of the magnetic field $\mathbf{B}$, and use the resistive gauge. For an isothermal gas the pressure is given by $p=\rho c_{s}^{2}$ , where $c_{s}$ is the sound speed, set to unity and $\rho$ is the density. The equations we solve are: 

\begin{equation}\label{1}
\frac{\partial\mathbf{A}}{\partial t}-\mathbf{v}\times\mathbf{B}-\eta\nabla^{2}\mathbf{A}=0,
\end{equation}

\begin{equation}
\frac{\text{D}\mathbf{v}}{\text{D}t}+c_{s}^{2}\nabla\ln\rho-\frac{1}{\rho}(\mathbf{j}\times\mathbf{B})-\frac{1}{\rho}\mathbf{F}_{\text{visc}}=0,
\end{equation}

\begin{equation}
\frac{\text{D}\ln\rho}{\text{D}t}+\nabla\cdot\mathbf{v}=0.
\end{equation}

Here $\mathbf{v}$ is the plasma fluid velocity, $t$ is time and the operator $\tfrac{\text{D}}{\text{D}t}=(\frac{\partial}{\partial t}+\mathbf{v}\cdot\nabla)$ indicates the material derivative. The viscous forces are given by $\mathbf{F}_{\text{visc}}=\nabla\cdot2\nu\rho\mathsf{S}$ with $\nu$ the kinematic viscosity and $\mathbf{\mathsf{S}}$ the traceless rate of strain tensor ($\mathbf{\mathsf{S}}=1/2(v_{i,j} + v_{j,i})-1/3\delta_{ij}\nabla\cdot \mathbf{v}$). The equations are solved on a grid size $256^3$. The gridsize of $256^3$ is more than adequate to resolve the spatial structures that we are interested in.  

\section{Open Boundary Conditions}
To assess the effect of the boundaries on the observed configuration we perform simulations using open boundary conditions in addition to the simulations with periodic boundary conditions reported in the main article. We simulated the $n=3, \ T=1.8$ initial condition using open boundary conditions. These conditions are implemented by imposing the condition that the fields (velocity $\mathbf{v}$ and magnetic field $\mathbf{B}$) are perpendicular to the boundary, allowing magnetic flux to exit the simulation volume, and a smooth variation of the density across the boundary by setting the first derivative across the boundaries to zero. 

\begin{figure} 
		\centering
			\includegraphics[width=0.48\textwidth]{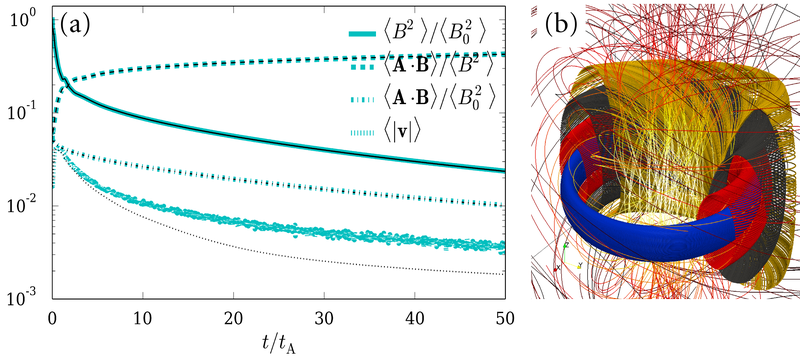}
	\caption{Comparison of periodic boundary conditions to open boundary conditions. (a) Quantities in the simulations decay in a similar fashion. The cyan fat curves are from the periodic boundary conditions simulation, whereas the thin black curves are from the open boundary conditions simulation. (b) Field structure in the open boundary conditions simulation at $t=42 t_{\rm A}$. The same nested structure appears.}
	\label{openboundary}
\end{figure}

Figure \ref{openboundary} shows the time evolution of relevant parameters. There is almost no difference between open boundaries and the periodic boundaries used, with the notable exception of the velocity field. The velocity field in the periodic boundaries is caused by the fluid motion in the pressure waves that are created in the initial collapse of the rings, but in the open boundary simulations these waves escape the volume. 

Also the magnetic fields are very similar in topology. The inset in figure \ref{openboundary} shows the magnetic field and toroidal surfaces, using the same seed points for the stream tracing as in figure 2 (a) of the paper.

\section{trefoil knot and single twisted ring}
\begin{figure} 
		\centering
			\includegraphics[width=0.48\textwidth]{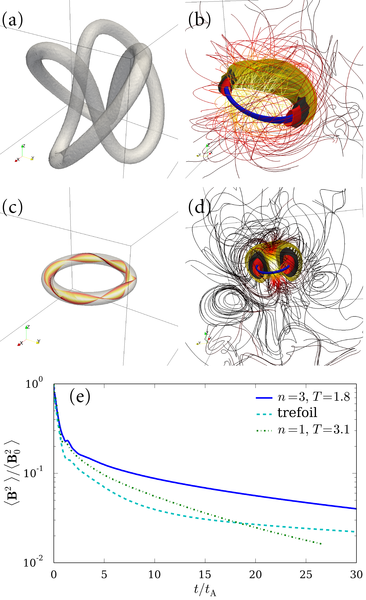}
	\caption{Initial conditions, final states, and evolution of magnetic energy for the trefoil and twisted torus simulation. (a), Initial state of the trefoil simulation. The field strength is 1 inside the tubes, and zero outside. (b), Final state of the trefoil simulation, showing the nested toroidal surfaces spanned by the field lines (clipped to show nesting) and 70 other field lines colored by magnetic intensity. (c) Initial condition for the twisted torus simulation. The field has a Gaussian profile. An isosurface of magnetic field strength 0.1 shows the approximate edge of the flux tube. The field strength is 1 in the center of the flux rings. A line of 100 closely spaced field lines shows the twist of the field. (d), Final configuration of the twisted torus simulation, visualized in the same way as (b). (e), The decay of magnetic energy for the $n=3,T=1$ simulation, the trefoil simulation, and the twisted torus simulation.}
	\label{torus_trefoil}
\end{figure}

In order to assess the generality of the observed emerging plasma configuration we also performed two other simulations with initial conditions that have helicity: The trefoil knot and a single twisted torus. The trefoil is parametrized as described in reference [22] in the main paper. The field is set to constant $1\mathcal{B}_0$ inside the rings and zero outside. The twisted torus is parametrized in the same way as the twisted rings, but the ring is set in the $x,y$-plane. The twist is set to $T=3.1$. The initial conditions can be seen in figure \ref{torus_trefoil} (a) and (c). All the simulation parameters are identical to the ones used in the paper: fully periodic simulation volume of $(2\pi l_0)$, isothermal gas with viscosity and magnetic diffusivity of $\eta= \nu = 2\times10^{-4}$. For easier comparison the same alfvenic timescale is used.

The fields evolve in the same manner as the linked fields rings in the paper, with an initial fast drop in magnetic energy, followed by a much slower decay. They also show the self-formed nested toroidal surfaces that appeared in the ring simulations, as is shown in figure \ref{torus_trefoil} (b) for the trefoil simulation (at $t=45t_{\rm A}$) and (d) for the twisted torus simulation (at $t=22.5t_{\rm A}$).   

\section{video}

The supplemental material contain a video visualizing the time evolution of the simulation with $n=3$ and $T=3.5$. This video shows the collapse of the rings and the emergence of the stable structure. Table \ref{moviestamps} shows at what times which processes are shown in the movie. 

\begin{table*}		
\caption{Video events and time, made from the simulation with $n=3$ and $T=3.5$}
	\centering
		\begin{tabular}{r|c|l}
		\hline\hline
			time (movie) & time (simulation) & event\\
			\hline
			0-5$s$   & 0-4.5$t_{\rm A}$   &initial collapse (slowed down)\\
			5-14$s$ & 4.5-34.7$t_{\rm A}$   &formation of structure \\
			14-17$s$  & 34.7$t_{\rm A}$     &plane showing $|\mathbf{B}|$ \\
			17-20$s$  & 34.7$t_{\rm A}$    & zoom into center of structure\\
			20-26$s$  & 34.7$t_{\rm A}$    & field line spanning first toroidal surface\\
			26-32$s$  & 34.7$t_{\rm A}$    & field line spanning second toroidal surface \\
			32-37$s$  & 34.7$t_{\rm A}$    & field line spanning third toroidal surface \\
			\hline
			\end{tabular}
	\label{moviestamps}
\end{table*}

\section{Time evolution of fields}

\begin{figure*}
	\centering
		\includegraphics[width=1\textwidth]{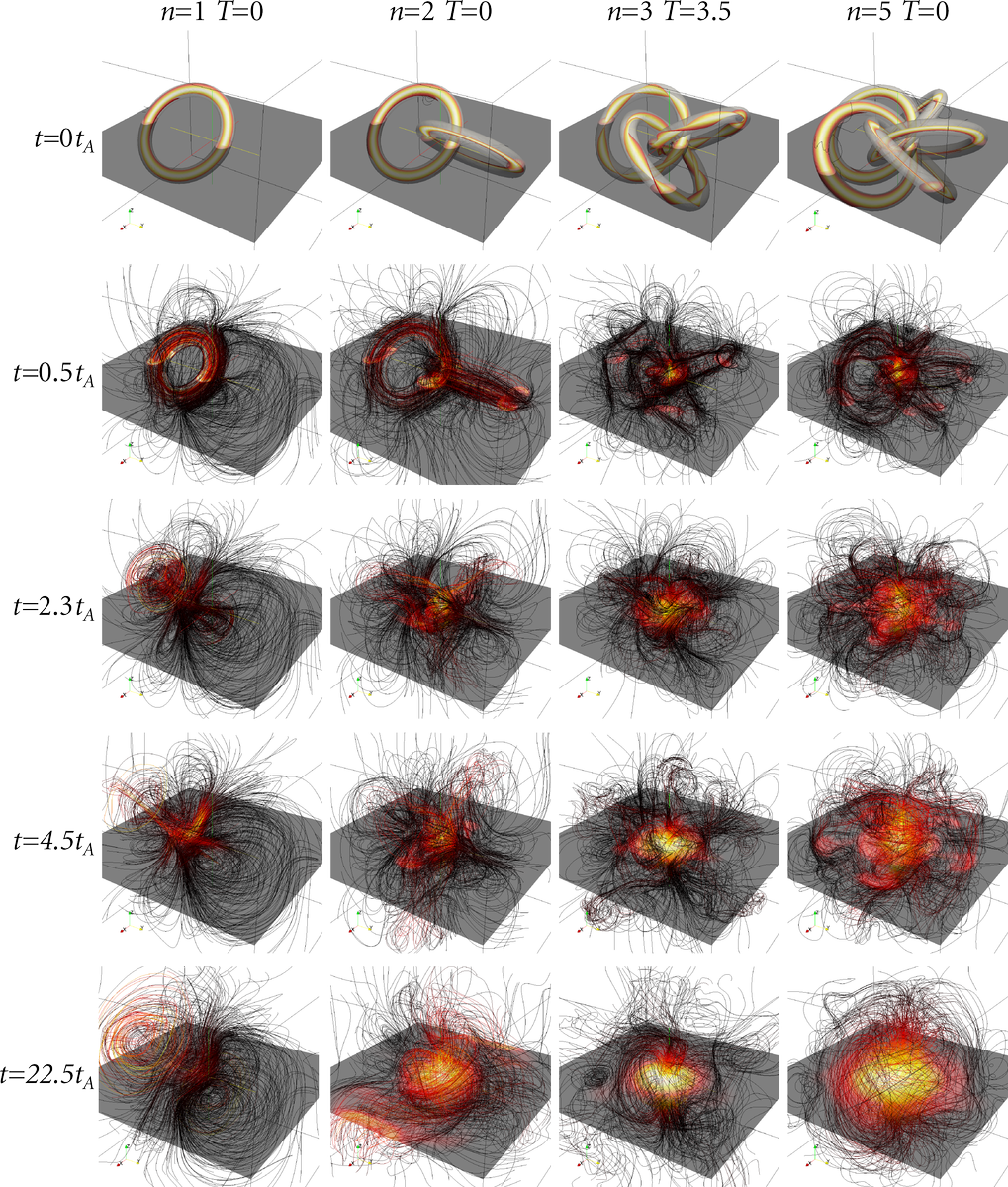}
	\caption{Time evolution of the magnetic fields. The properties of the initial condition are given on the top of each column. Time increases in the downward direction. In the initial condition an isosurface for the magnetic energy is drawn indicating the extent of the ring. The configurations with more than one ring show the emergence of a localized linked field configuration, the single ring does not. At $t=22.5t_{\rm A}$ details of the initial structure are lost and a universal structure of linked field lines on nested toroidal surfaces appears.}
	\label{largefig}
\end{figure*}

Figure \ref{largefig} shows in more detail the time evolution of different initial conditions, including the evolution of just a single ring. The first row shows the initial condition. In the second row, taken at $t=0.5t_{\rm A}$ it is seen that the rings have contracted. This is to be expected, as this lowers the magnetic energy. Another way of understanding this is by looking at the currents that must source this field. A flux tube is sourced by ring-currents that run around the minor radius of the tube. These currents run parallel, and therefore must attract one another. The net effect is a force that causes the magnetic flux tube to contract. 

If one looks closely at the difference between the $n=1$ and $n=2$ situation one can see the physical manifestation of helicity conservation in action. The single ring can contract onto itself, but the two rings cannot cross each other. 

When the rings have collapsed, shown in the third row, the single ring can then lose a large amount of field, that is carried away from the center by fluid motion. This cannot happen in the linked rings. 

The fourth row shows the state at time $t=4.5t_{\rm A}$. It is clear that in the linked configuration the energy is still confined in the center, but the initial structure is still seen. At time $22.5t_{\rm A}$ most of the initial structure is lost, and only a localized, helical configuration is seen for the linked configurations, whereas little structure is visible in the simulation with a single ring. 

\section{Calculation of the magnetic field}
Consider an adaptation of the Hopf map 
\begin{equation}
h^{(\omega_1,\omega_2)}:= \pi^{(2)^{-1}}\left(\frac{z_1^{(\omega_2)}}{z_2^{(\omega_1)}}\right).
\end{equation}
Here $z^{(w)}$ indicates the operation on a complex number such that only its argument is multiplied by the real number $w$. The two complex numbers $(z_1,z_2)\subset \mathbb{C}^2$ with $|z_1^2|+|z_2^2|=1$ define a point on the three-sphere $S^3$ and $\pi^{(2)^{-1}}:\mathbb{C}\rightarrow S^2$ is inverse stereographic projection from the north pole. From this definition follows that $h^{(\omega_1,\omega_2)}(z_1,z_2)=h^{(\omega_1,\omega_2)}(e^{i\omega_1\theta}z_1,e^{i\omega_2\theta}z_2)$ for all values of $\theta$. Consequently the pre-image of a point on $S^2$ is a continuous curve in $S^3$. Furthermore, if $\omega_1=\omega_2$ each curve is a great circle in $S^3$, and the map is in essence identical to the Hopf map. For $\omega_1\neq\omega_2$ the pre-image of a point in $S^2$ is a curve in $S^3$ that oscillates in the $z_1$-direction with frequency $\omega_1$ and in the $z_2$-direction with frequency $\omega_2$. 

To construct a field in $\mathbb{R}^3$ define the following map: 
\begin{equation}
\phi:\mathbb{R}^3 \rightarrow \mathbb{C}, ~\phi= \pi^{(2)} \circ h^{(\omega_1,\omega_2)} \circ \pi^{(3)^{-1}},
\end{equation}
with $\pi^{(3)^{-1}}:\mathbb{R}^3\rightarrow S^3$ inverse stereographic projection (also from the north pole of $S^3$) to the three-sphere. This map has the following explicit form:
\begin{equation}
\phi(x, y, z)=\frac{\left(  2(x+ i y) \right)^{(\omega_2)} }{\left(  2z + i(r^2-1) \right)^{(\omega_1)} },
\label{Phi}
\end{equation}
where $r^2 = x^2 + y^2 + z^2$.  We get a vector field in $\mathbb{R}^3$ by calculating
\begin{equation}
\mathbf{B} = \frac{\sqrt{a}}{2\pi i}\frac{\nabla\phi \times \nabla \phi^*}{(1 + \phi\phi^*)^2}.
\label{BfromPhi}
\end{equation}
Here $\sqrt{a}$ is a constant such that the magnetic field has correct dimensions. The cross product between $\nabla\phi$ and $\nabla\phi^*$ is a vector in the direction that $\phi$ remains constant. The curves of constant $\phi$ are continuous, oscillating and closed (or, for incommensurate $\omega_1$ and $\omega_2$, dense in a compact subspace of $S^3$) curves in $S^3$, and will thus be  so too in $\mathbb{R}^3\cup\infty$. We calculate this field, and then scale it with a factor $r_0$ by substituting $\left( x,y,z \right)  \rightarrow \left(  \tfrac{x}{r_0},\tfrac{y}{r_0},\tfrac{z}{r_0} \right)$ to obtain the expression for a magnetic field:
\begin{equation}
\mathbf{B} = \frac{4r_0^4\sqrt{a}}{\pi(r_0^2 + r^2)^3}
\begin{pmatrix} 2(\omega_2 r_0 y- \omega_1 xz  ) \\ -2( \omega_2 r_0 x + \omega_1 yz) \\ \omega_1(-r_0^2+x^2+y^2-z^2)  \end{pmatrix}.
\label{Bfield}
\end{equation}

This field is cylindrically symmetric as can be seen by the absence of a $\phi$-dependence if equation \ref{Bfield} is put in cylindrical coordinates:

\begin{multline}
\mathbf{B}(r,\phi,z)=\frac{4r_0^2\sqrt{a}}{\pi\left(r_0^2 + r^2 + z^2 \right)} (-2\omega_2r_0r\hat{\phi} \\ - 2\omega_1 z r \hat{r} + \omega_1(-r_0^2+r^2-z^2) \hat{z} ).
\label{cylfield}
\end{multline}

Every field line lies on the surface of a member of a set of nested tori. The smallest reduces to a circle with radius of $r_0$ (magnetic axis), and the largest is a line along the $z$-axis. The field lines wind around the poloidal direction with frequency $\omega_1$, and toroidal direction with frequency $\omega_2$. If $\frac{\omega_1}{\omega_2}$ is rational, $\frac{\omega_1}{\omega_2}=\frac{n}{m}$, and every field line is a $(n,m)$ torus knot. We stress that every integral curve of this field is itself a knot (or circle, or ergodically spanning the toroidal surface), but the global field is smooth and continuous. The ratio $\frac{\omega_1}{\omega_2}$ gives the ratio of toroidal to poloidal winding of the curves or the rotational transform $\imath$.

\section{Calculating the helicity and magnetic energy of the Sagdeev field}

The vector potential corresponding with the field in equation \ref{Bfield}, defined as $\nabla\times\mathbf{A}=\mathbf{B}$, is given by

\begin{equation}
\mathbf{A} = \frac{r_0^3\sqrt{a}}{\pi(r_0^2 + r^2)^2}
\begin{pmatrix} 2(r_0\omega_1 y- \omega_2 xz  ) \\ -2( r_0 \omega_1 x +\omega_2 yz) \\ \omega_2(-r_0^2+x^2+y^2-z^2)  \end{pmatrix}.
\label{Afield}
\end{equation}

The inner product $\mathbf{A}\cdot\mathbf{B}$ is given by 
\begin{equation}
\mathbf{A}\cdot\mathbf{B}=\frac{4ar_0^7\omega_1\omega_2}{\pi^2(r_0^2+r^2)^3}.
\end{equation}
The helicity is then
\begin{equation}\label{helicity}
H_{\rm m}=\int \frac{4 a r_0^7\omega_1\omega_2}{\pi^2(r_0^2+r^2)^3} \ \mathrm{d}^3x=a r_0^4\omega_1\omega_2.
\end{equation}

The value of $B^2$ is given by
\begin{equation}
\mathbf{B}\cdot\mathbf{B}=\frac{16a r_0^8}{\pi^2(r_0^2+r^2)^6}\left(4r_0^2(\omega_2^2-\omega_1^2)(x^2+y^2)+\omega_1^2(r_0^2+r^2)^2\right),
\end{equation}
which allows us to calculate the integral of $B^2$ over all space
\begin{equation}\label{Bint}
\int B^2 \ \mathrm{d}^3x=a r_0^3(\omega_1^2+\omega_2^2).
\end{equation}

\section{Finding the smallest invariant torus and orientation of the magnetic structure}

\begin{figure}
		\centering
			\includegraphics[width=0.48\textwidth]{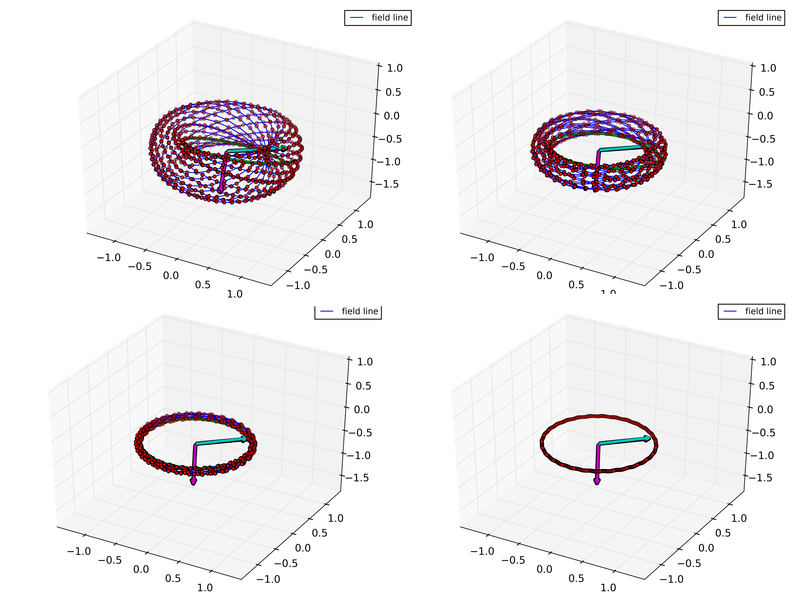}
	\caption{fitting routine for finding the smallest torus. The magenta arrow indicates the vector $\mathbf{\hat{n}}$, and the cyan arrow indicates the vector $r_0\mathbf{\hat{p}}$, which is the seed point for the next iteration in the fitting routine for the torus.}
	\label{convergence}
\end{figure}

Properties of the field structure are extracted from the simulations. First the radius of the magnetic axis is found by a fitting routine. 

A point near the magnetic axis is guessed. from that point a the field is integrated using a fixed-step size field integration routine for a distance of $500l_0$. A set of points ${\mathbf{x_i}}$ on this field line is returned. The geometrical center of the torus is easily calculated by the weighted average of all the points:
\begin{equation}
\mathbf{x}_0=\langle\mathbf{x}_i\rangle.
\end{equation}
An approximate major radius of the torus can be found by calculating
\begin{equation}
r_0=\langle \mathbf{x}_i-\mathbf{x}_0 \rangle.
\end{equation}
To find the orientation of the torus we calculate the cross product between the vector from the center to a point on the torus ($\mathbf{x}_i-\mathbf{x}_0$) and a vector from a point on the torus to the next point ($\mathbf{x}_{i+1}-\mathbf{x}_i$) to get a vector pointing out of the torus and average this over all points on the field line:
\begin{equation}
\mathbf{\hat{n}}= \frac{\langle(\mathbf{x}_i-\mathbf{x}_0)\times (\mathbf{x}_{i+1}-\mathbf{x}_i)\rangle}{|(\mathbf{x}_i-\mathbf{x}_0)\times (\mathbf{x}_{i+1}-\mathbf{x}_i)|} .
\end{equation} 
A vector perpendicular to the normal vector is found  by solving 
\begin{equation}
\mathbf{\hat{n}} \cdot \begin{pmatrix} 1 \\ 1 \\ c  \end{pmatrix}=0
\end{equation}
for $c$ and defining the vector
\begin{equation}
\hat{\mathbf{p}}=\frac{1}{\sqrt{2+c^2}}\begin{pmatrix} 1 \\ 1 \\ c  \end{pmatrix}.
\end{equation} 
This yields a perpendicular vector in all cases except if the normal vector is exactly in the $\begin{pmatrix} 1\\1\\0  \end{pmatrix}$ direction. 

Using the properties calculated above, a new starting point is taken to be
\begin{equation}
\mathbf{s}=\mathbf{x}_0+r_0\mathbf{\hat{p}}.
\end{equation}
This starting point lies inside of the volume spanned by the previous torus. Integrating a new field line from this point yields a new field torus, but since the starting point is inside the previous torus this torus is nested inside the previous.  
This routine is repeated until the difference between two subsequent starting points reaches below a threshold of $0.005l_0$. 

Figure \ref{convergence} shows the first few steps of this iterative process, with $\mathbf{\hat{n}}$ and $\mathbf{s}$ starting from $\mathbf{x}_0$ drawn in magenta respectively in cyan. These are the properties extracted from the simulation: the radius of the smallest invariant torus given by $|\mathbf{s} -\mathbf{x}_0|$, the geometrical center of the invariant torus $\mathbf{x}_0$ and it's orientation given by $\mathbf{\hat{n}}$

\section{Finding the rotational transform of a surface and parameters for analytical expression}

The rotational transform of field lines spanning invariant tori is found in the following way: First we calculate the number of times the field line crosses the plane defined by $\mathbf{x}_0$ and $\mathbf{\hat{n}}$. This number is twice the the toroidal winding $n_{\rm p}$, and is found by calculating the distance of each point in the curve to this plane and counting the zero crossings. The number of times the curve crosses the plane defined by $\mathbf{x}_0$ and $\mathbf{\hat{p}}$ is twice the toroidal winding $n_{\rm t}$. For a sufficiently long field line the ratio of poloidal to toroidal winding approaches the rotational transform $n_{\rm p}/n_{\rm_t} \simeq \imath$.

We describe how to calculate the variables $\omega_1$, $\omega_2$ and $r_0$ of equation \ref{Bfield} of the paper, that describe a field with the same helicity, magnetic energy, and rotational transform of the field lines as the simulation field. In the dimensionless units of our simulations we can take the constant $\sqrt{a}$ to be equal to 1.

To find the ratio of toroidal winding to poloidal winding, we calculate the rotational transform of a single field line with a starting point $\mathbf{s}_T=\mathbf{x}_0+1.15r_0\mathbf{\hat{p}}$ and a length of $1000l_0$. The rotational transform is roughly constant on the magnetic surfaces, so it is sufficient to extract it only from a single curve. The rotational transform $\imath$ is then equal to the ratio of poloidal to toroidal winding, and can be used to set the ratio of$\omega_1/\omega_2$. 

The value of $r_0$ is the radius of the smallest of the nested tori, and is found by using the fitting routine described above. The value of $\int B^2 \ \mathrm{d^3x}$ is calculated for the entire field in the simulation, and that value is used as the answer for equation \ref{Bint}. A field approximating the output of the simulation will have values of $\omega_1$ and $\omega_2$ that are then uniquely defined through the ratio $\frac{\omega_1}{\omega_2}$ and equation \ref{Bint}, and the sign of the total helicity in equation 3 of the paper. The results for all simulations are given in table \ref{fieldprops}.

\begin{table} \caption{Field Properties} 
	\centering 
 		 \begin{tabular}{c | c| c| c| c| c| c| c}
 		\hline\hline
			 n & T & t& $r_0$ & $\omega_1$ & $\omega_2$ & $r_0^3(\omega_1^2+\omega_2^2 )$ &$\tfrac{\omega_1}{\omega_2}$\\ 
			 \hline 			2&0 & -  &  -    &   -    &   -    &   -    &   -   \\ 
			 \color{gray} 3&\color{gray}0 &$54.0t_A$    &\color{gray} 1.25 &\color{gray} 0.049 & \color{gray}0.090 &\color{gray} 0.020 &\color{gray} 0.54 \\ 
			 4&0   &$22.5t_A$   & 1.10 & 0.17 & 0.19 & 0.085 & 0.90 \\ 
			 5&0   &$22.5t_A$  & 1.13 & 0.22 & 0.23 & 0.141 & 0.96 \\ 
			 6&0   &$22.5t_A$   & 1.14 & 0.26 & 0.25 & 0.191 & 1.05 \\ 
			\hline
			 3&0.9 &$22.5t_A$  & 0.90 & 0.18 & 0.19 & 0.048 & 0.95 \\ 
			 3&1.8 &$22.5t_A$  & 0.78 & 0.24 & 0.27 & 0.063 & 0.89 \\ 
			 3&2.6 &$22.5t_A$  & 0.74 & 0.30 & 0.32 & 0.075 & 0.92 \\ 
			 3&3.5 &$22.5t_A$  & 0.75 & 0.32 & 0.33 & 0.086 & 0.98 \\ 
			 3&4.4 &$22.5t_A$  & 0.80 & 0.32 & 0.30 & 0.096 & 1.05 \\ 
			 \hline 
 		\end{tabular} 
 \label{fieldprops} 
\end{table}

The simulation with $n=2$ and $T=0$ did not become ordered enough to be analyzed with the described routine before the simulation was stopped at $t=60t_{\rm A}$, the simulation with $n=3$ and $T=0$ only at $t=54t_{\rm A}$. This explains the different value for $r_0$ and much smaller value for $\omega_1/\omega_2$.  All simulations show the emergence of a configuration where $r_0$ is  roughly 1, and $r_0$ becomes smaller if $T$ is larger. The value of $\omega_1/\omega_2$ is also around one, and increases with higher initial helicity. 
\end{document}